\documentclass[twocolumn]{aastex631}

\begin{document}

\title{Our Halo of Ice and Fire:\\Strong Kinematic Asymmetries in the Galactic Halo}

\author[0000-0002-6800-5778]{Jiwon Jesse Han}
\affiliation{Center for Astrophysics $|$ Harvard \& Smithsonian, 60 Garden Street, Cambridge, MA 02138, USA}

\author[0000-0002-1590-8551]{Charlie Conroy}
\affiliation{Center for Astrophysics $|$ Harvard \& Smithsonian, 60 Garden Street, Cambridge, MA 02138, USA}

\author[0000-0002-5177-727X]{Dennis Zaritsky}
\affiliation{Steward Observatory, University of Arizona, 933 North Cherry Avenue, Tucson, AZ 85721-0065, USA}

\author[0000-0002-7846-9787]{Ana Bonaca}
\affiliation{The Observatories of the Carnegie Institution for Science, 813 Santa Barbara St., Pasadena, CA 91101, USA}

\author{Nelson Caldwell}
\affiliation{Center for Astrophysics $|$ Harvard \& Smithsonian, 60 Garden Street, Cambridge, MA 02138, USA}

\author[0000-0002-0572-8012]{Vedant Chandra}
\affiliation{Center for Astrophysics $|$ Harvard \& Smithsonian, 60 Garden Street, Cambridge, MA 02138, USA}

\author[0000-0001-5082-9536]{Yuan-Sen Ting}
\affiliation{Research School of Astronomy \& Astrophysics, Australian National University, Cotter Road, Weston Creek, ACT 2611, Canberra, Australia}
\affiliation{School of Computing, Australian National University, Acton ACT 2601, Australia}
\affiliation{Department of Astronomy, The Ohio State University, Columbus, OH 43210, USA}
\affiliation{Center for Cosmology and AstroParticle Physics (CCAPP), The Ohio State University, Columbus, OH 43210, USA}

\begin{abstract}

The kinematics of the stellar halo hold important clues to the assembly history and mass distribution of the Galaxy. In this study, we map the kinematics of stars across the Galactic halo with the H3 Survey. We find a complex distribution that breaks both azimuthal symmetry about the $Z$-axis and mirror symmetry about the Galactic plane. This asymmetry manifests as large variations in the radial velocity dispersion $\sigma_r$ from as ``cold'' as 70 $\text{km}\text{ s}^{-1}$ to as ``hot'' as 160 $\text{km}\text{ s}^{-1}$. We use stellar chemistry to distinguish accreted stars from in-situ stars in the halo, and find that the accreted population has higher $\sigma_r$ and radially biased orbits, while the in-situ population has lower $\sigma_r$ and isotropic orbits. As a result, the Galactic halo kinematics are highly heterogeneous and poorly approximated as being spherical or axisymmetric. We measure radial profiles of $\sigma_r$ and the anisotropy parameter $\beta$ over Galactocentric radii $10-80\text{ kpc}$, and find that discrepancies in the literature are due to the nonspherical geometry and heterogeneous nature of the halo. Investigating the effect of strongly asymmetric $\sigma_r$ and $\beta$ on equilibrium models is a path forward to accurately constraining the Galactic gravitational field, including its total mass.

\end{abstract}

\section{Introduction}

The stellar halo of the Galaxy possesses a duality. On the one hand, long relaxation times in the halo allow for stars to keep a ``fossil record'' of the hierarchical formation history of the Galaxy; this is the subject of Galactic archaeology \citep{els62, searle78, freeman02}. On the other hand, the stellar halo reflects the equilibrium kinematics of the underlying dark matter halo \citep{BT87}. Past works have assumed equilibrium to apply the Jeans equations \citep{jeans15} to constrain global properties of the Galaxy, notably its total mass \citep[e.g.,][]{hartwick78, battaglia05, dehnen06, gnedin10, deason12}{}. Such modeling efforts involve two steps. First, the halo velocity ellipsoid is observationally determined, often parameterized by the Galactocentric radial velocity dispersion $\sigma_r$ and the anisotropy parameter $\beta\equiv 1 - \frac{\sigma_\phi^2+\sigma_\theta^2}{2\sigma_r^2}$, where $\sigma_\phi$ and $\sigma_\theta$ are Galactocentric longitudinal and latitudinal velocity dispersion. Second, the Jeans equations are solved to yield the total enclosed mass $M(<r)$. In both steps, the assumption of axisymmetry plays a critical role. Observationally, kinematic tracers such as halo stars or globular clusters (excluding unrelaxed structures such as the Sagittarius stream) are spherically averaged in order to measure $\sigma_r$ and $\beta$ over a certain Galactocentric radial range \citep[e.g.,][]{gnedin10,watkins19, lancaster19, bird22}{}{}. Theoretically, the Jeans equations are solved in either 1D (spherical $r$) or 2D \citep[cylindrical $R$ and $Z$, e.g.,][]{cappellari08} coordinates that assume symmetry about the $Z$ axis. The results from applying this analysis to the Galaxy have varied considerably, producing up to a factor of two discrepancy in the mass of the Milky Way just from Jeans modeling alone \citep[see, e.g.,][for a review]{wang20}. This discrepancy lies at the heart of the duality of the stellar halo: how does the fossil record affect the accuracy of equilibrium modeling?

\begin{figure}
    \centering
    \includegraphics[width=0.48\textwidth]{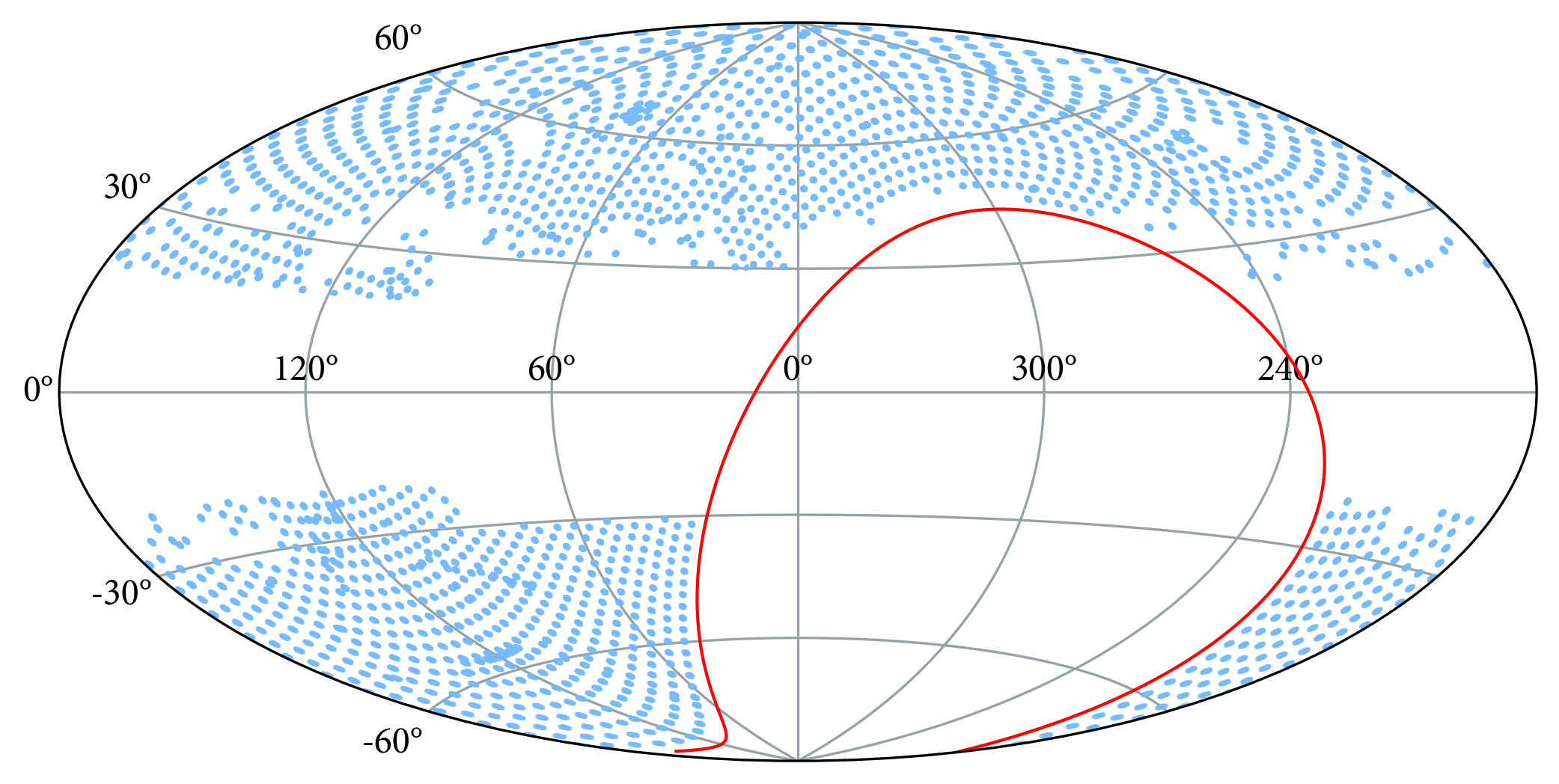}
    \caption{Footprint of the H3 Survey as of April 2024 in Galactic $l,b$ coordinates. Each blue circle indicates a $1^{\circ}$ diameter field of view. The survey does not reach the Galactic plane ($|b|<20^{\circ}$) or the Southern Hemisphere ($\delta<-20^{\circ}$, marked in red line).}
    \label{fig:foot}
\end{figure}

From \textit{Gaia} DR2 \citep{gaia18}, studies have found that a single radial merger $8-10\text{ Gyr}$ ago has played a dominant role in creating the stellar halo of the Galaxy \citep{belokurov18, helmi18}. The remnant of this merger has been dubbed \textit{Gaia}-Sausage-Enceladus (GSE). The degree to which GSE is dynamically relaxed at present day is under debate. Some studies find that $8-10\text{ Gyr}$ is enough time for GSE to become completely spherical, particularly if one assumes a spherical dark matter halo \citep{balbinot21}. Others find observational evidence for a non-spherical GSE at present day that is triaxial \citep{iorio19} and tilted with respect to the Galactic disk \citep{han22b}. The misalignment of the disk and the stellar halo has been shown to imply a dark matter halo that is tilted in a similar direction (hence, necessarily nonspherical) in both idealized \citep{han22} and cosmological simulations \citep{han23b}. More generally, cosmological simulations produce a wealth of present-day non-spherical halos that are shaped by previous mergers and the larger cosmological environment, often changing shapes and direction sharply as a function of radius \citep[e.g.,][]{prada19, shao21, emami21}.

If indeed the bulk of the Galactic stellar halo exhibits triaxiality and misalignment with the disk, these asymmetries will also manifest in the kinematics of halo stars. Strong asymmetries in tracer kinematics would have implications for both the validity of spherical Jeans modeling and the spherically averaged measurements of $\sigma_r$ and $\beta$. Here, we use the H3 survey to map the kinematics of stars across a wide region in the Galactic halo. We then use this map to directly evaluate the spherical symmetry of $\sigma_r$ and $\beta$.

The paper is organized as follows. In Section \ref{sec:methods} we introduce the H3 survey and the halo sample used in this study. In Section \ref{sec:results} we present a Galactocentric map of $\sigma_r$, separating the accreted and in-situ halo based on chemistry. We also show the same map in Galactic $l,b$ projection. We then compare our measurements to the literature by computing the radial profile for $\sigma_r$ and $\beta$. We close by discussing the implications of these results in Section \ref{sec:discussion}.

\begin{figure}
    \centering
    \includegraphics[width=0.5\textwidth]{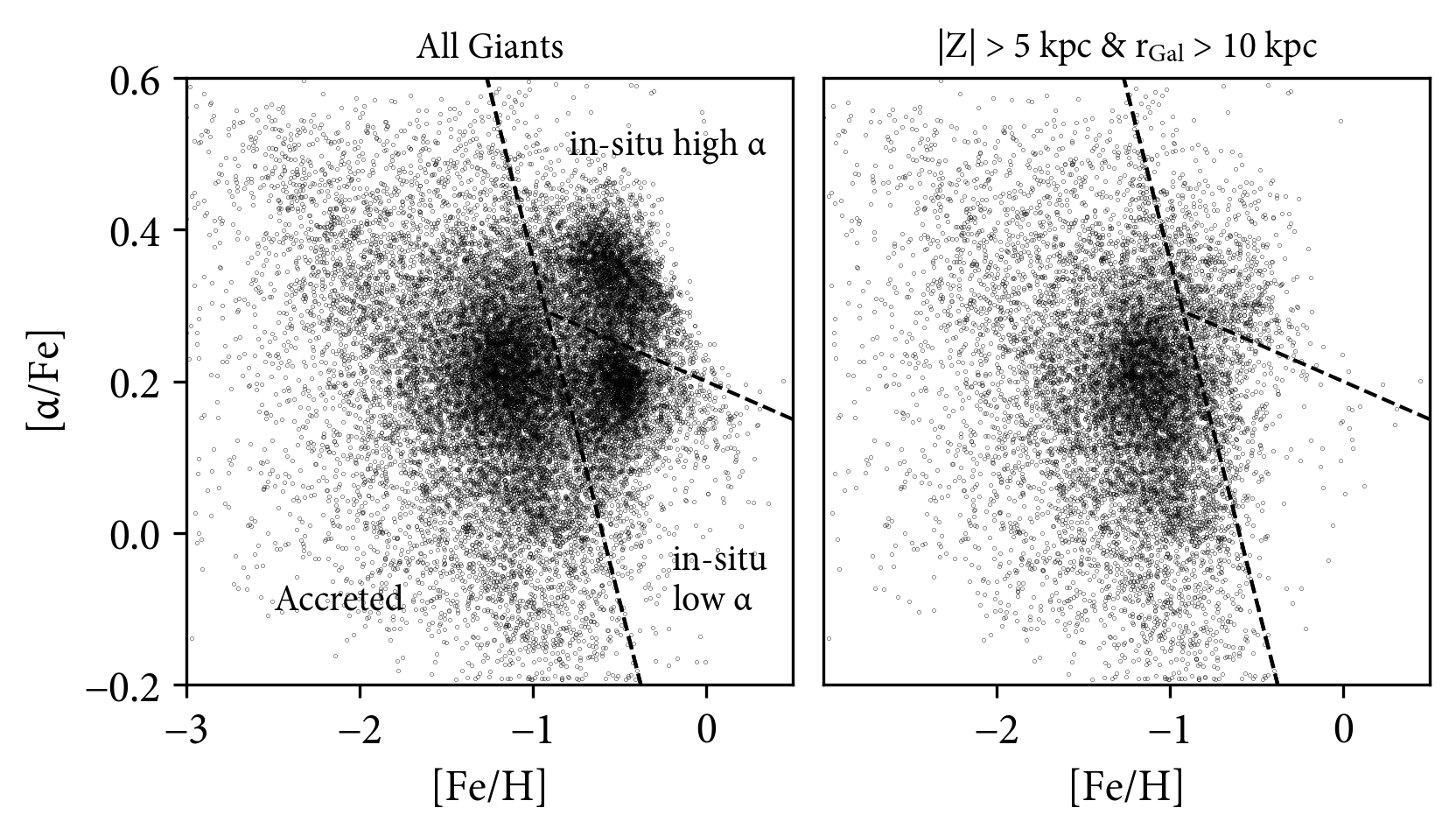}
    \caption{Chemical selection of H3 giants in the [Fe/H]-[$\alpha$/Fe] plane. The left panel includes all giants that are not in cold substructures. There are three notable overdensities: the low-[Fe/H] accreted stars, the high-[$\alpha$/Fe] in-situ stars (``thick disk'' chemistry), and the low-[$\alpha$/Fe] in-situ stars (``thin disk'' chemistry). Among the low-[$\alpha$/Fe] stars, those that are on prograde, circular orbits (eccentricity less than 0.3) are attributed to Aleph and removed from the halo sample. The right panel shows the final sample used in this study, which does not distinguish the two $\alpha$ sequences within the in-situ sample.}
    \label{fig:chem_sel}
\end{figure}

\section{Data \& Methods}\label{sec:methods}

H3 is a high Galactic latitude spectroscopic survey designed to target halo stars \citep{conroy19}. With a relatively simple target selection based on magnitude ($r<18$), sky position ($|b|>20^{\circ}$ and $\delta>-20^{\circ}$), and \textit{Gaia} parallax ($\pi<0.5\text{ mas}$), H3 offers a wide and deep view of the halo. We show an up-to-date footprint of the survey in Figure \ref{fig:foot}. By combining $R\sim32,000$ spectra in the wavelength range $515\text{nm}-530\text{nm}$ with optical to near-infrared photometry, H3 measures radial velocities and stellar parameters (effective temperature, surface gravity, metallicity, and $\alpha$-element abundance) along with isochrone distances using the \texttt{MINESweeper} code \citep{cargile20}. These distances can then be used to convert radial velocities and \textit{Gaia} proper motions into Galactocentric 3D positions and 3D velocities. As of Dec 2023, H3 has collected 302,485 stars, 33,068 (12,167) of which are $>$ 5 kpc (10 kpc) away from the Sun and have good measurements (signal-to-noise ratio per pixel greater than 3, and have a successful fit from \texttt{MINESweeper}). For a more thorough review of the H3 Survey, we direct the reader to \cite{conroy19}.

\begin{figure}
    \centering
    \includegraphics[width=0.43\textwidth]{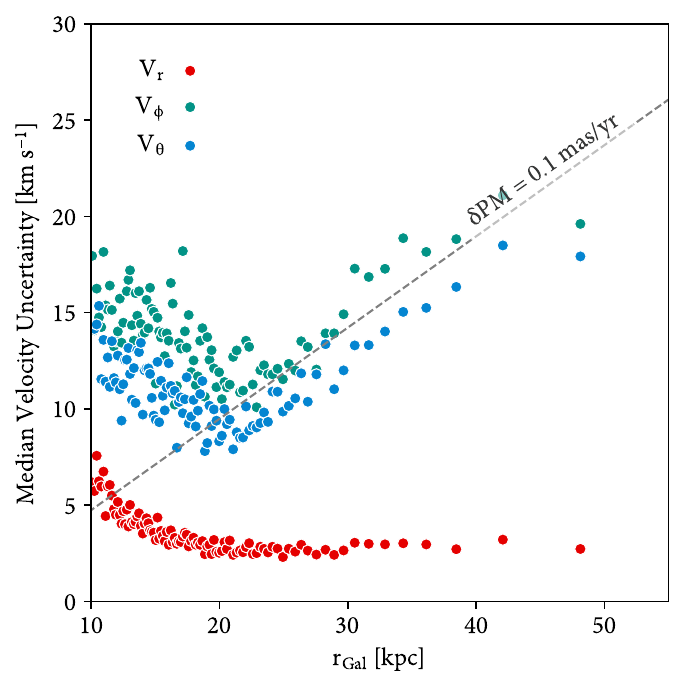}
    \caption{Galactocentric radial (red) and tangential (green, blue) velocity uncertainties as a function of Galactocentric radius. Each circle represents the median velocity uncertainty in radial bins containing $[n,n+1)^{th}$ percentile of the sample, where $n \in \{1,2,... 99 \}$. The grey dashed line indicates the tangential velocity uncertainties of a constant 0.1 $\text{mas}\text{ yr}^{-1}$ proper motion uncertainty at a distance of $r\text{ kpc}$. At large radii, the tangential velocity uncertainties increase linearly along this line, while the radial velocity uncertainties are roughly constant at $\sim2\text{ km}\text{s}^{-1}$.}
    \label{fig:1}
\end{figure}

\begin{figure*}
    \centering
    \includegraphics[width=\textwidth]{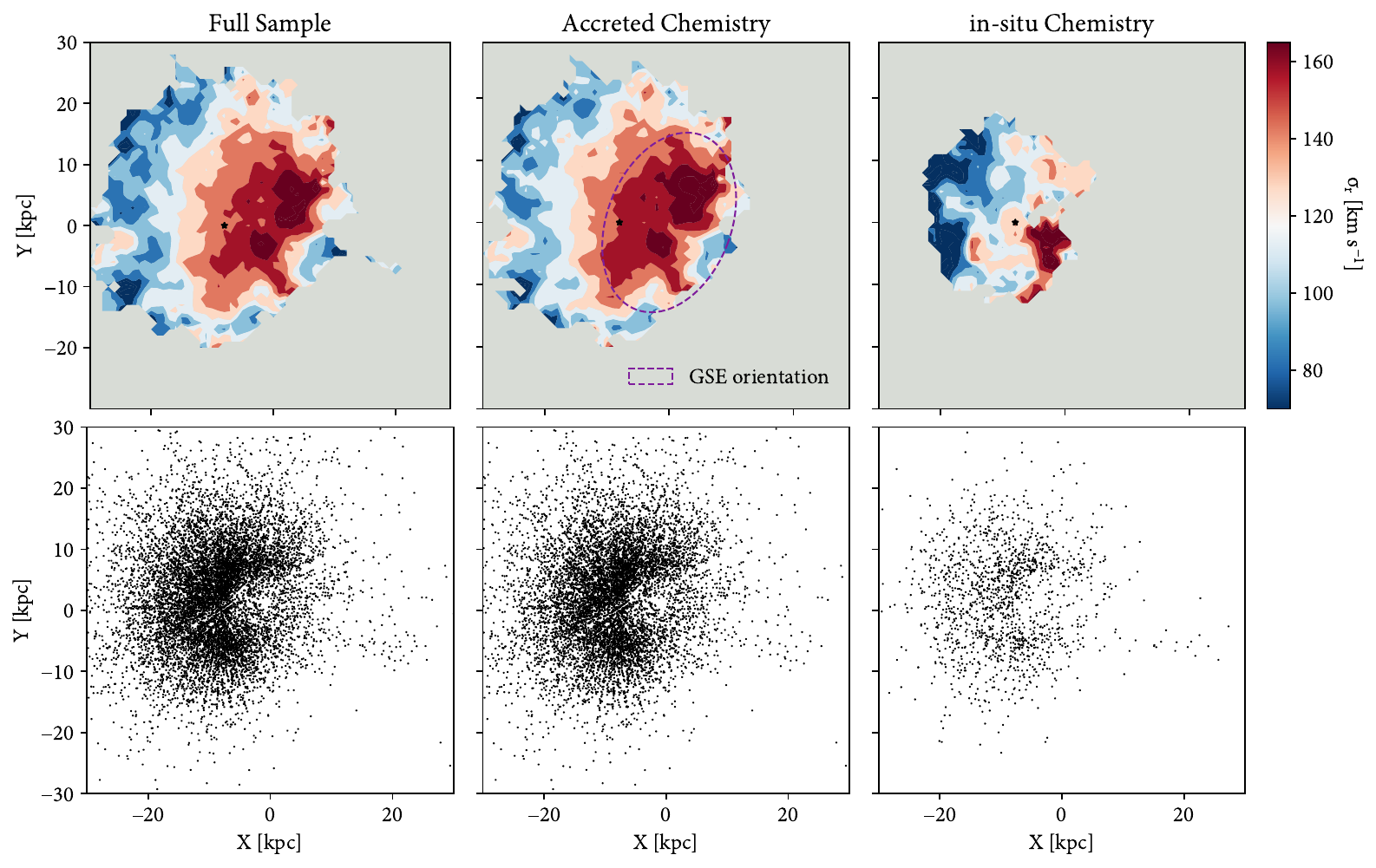}
    \caption{Galactocentric $XY$ projection of $\sigma_r$, with the solar location marked as a black star. We construct a grid from -30 to +30 kpc in X and Y in increments of 1 kpc. At each grid point, we measure $\sigma_r$ of halo stars (as defined geometrically in Section \ref{sec:methods}) within a 2 kpc radius. The resulting grid of $\sigma_r$ values are visualized as a contour plot with 11 levels. We perform this analysis for three samples: the whole halo sample (left panel), the accreted halo (center panel), and the in-situ halo (right panel). For the accreted sample, we overplot the iso-density contour of \textit{Gaia}-Sausage-Enceladus (GSE) as measured by \cite{iorio19} in purple dashed ellipse. The accreted $\sigma_r$ contours display a strong azimuthal anisotropy which aligns with the orientation of GSE. In the bottom panels, we show the individual data points in each sample.}
    \label{fig:xy}
\end{figure*}

For the purpose of this study, we limit our sample to giant stars based on $\log(g)<3.5$ and exclude kinematically cold structures such as dwarf galaxies, globular clusters, and the Sagittarius stream. We identify Sagittarius stream member stars based on chemistry and angular momenta $L_z$ and $L_y$, as described in \cite{Johnson20}. The completeness of this selection is very high, and we expect minimal contamination from Sagittarius in the sample after this cut, even in fields that are not on the bulk of the stream. Furthermore, we exclude Aleph, a halo substructure towards the Galactic anticenter of yet unknown origin \citep{naidu20}, characterized by its disk-like chemistry and highly circular orbits. Aleph is likely a high-latitude extension of the Monoceros Ring and/or the flared stellar disk \citep{momany06}, and will be a topic of future study. Lastly, we exclude stars that are on unbound orbits based on non-negative orbital energies computed from a model Milky Way potential \citep{bovy15, price-whelan17}.

Once we have removed the unrelaxed substructures and unbound stars, we do not apply additional kinematic criteria that could directly affect the velocity distribution of the sample. Instead, the remaining halo sample is selected purely geometrically based on Galactocentric $|Z|>5\text{ kpc}$ and spherical $r_{\text{Gal}}>10\text{ kpc}$. This selection avoids the thick disk of the Milky Way by more than five scale heights \citep{bland-hawthorn16}. We note that the disk of the Galaxy warps towards large radii, but the amplitude of the stellar warp is less than 2 kpc at cylindrical $R_{\text{Gal}}=20\text{ kpc}$ \citep{chen19}. Hence, we expect our $|Z|$ cut to exclude most of the stars on the disk warp, although some contamination is still possible. The geometric cut results in a total of 10,469 halo stars. 

A key feature of the H3 survey is the [$\alpha$/Fe] measurement in addition to [Fe/H]. The 2D chemistry information enables a clean separation of the in-situ component halo from the accreted components, since distinct stellar populations follow unique sequences in the [Fe/H]---[$\alpha$/Fe] plane \citep{tinsley79}. This chemical separation is crucial to this study, since it allows us to investigate the origin of halo stars without biasing $\sigma_r$ or $\beta$ measurements. We adopt the [$\alpha$/Fe]---[Fe/H] selection similar to \cite{han22b} that separates the accreted halo from the in-situ halo, as shown in Figure \ref{fig:chem_sel}. In the left panel we show all of the giants in H3 that are not in cold substructures, revealing three major overdensities: the accreted stars at low [Fe/H], the high [$\alpha$/Fe] in-situ sequence (``thick disk'' chemistry), and the low [$\alpha$/Fe] in-situ sequence (``thin disk'' chemistry). While we do not distinguish the in-situ low-$\alpha$ and high-$\alpha$ sequence here, we do use this information to remove Aleph, which is a low-$\alpha$ substructure. In the right panel we show the geometrically selected halo sample. The final halo sample comprises 9353 accreted stars and 1116 in-situ stars.

\begin{figure*}
    \centering
    \includegraphics[width=\textwidth]{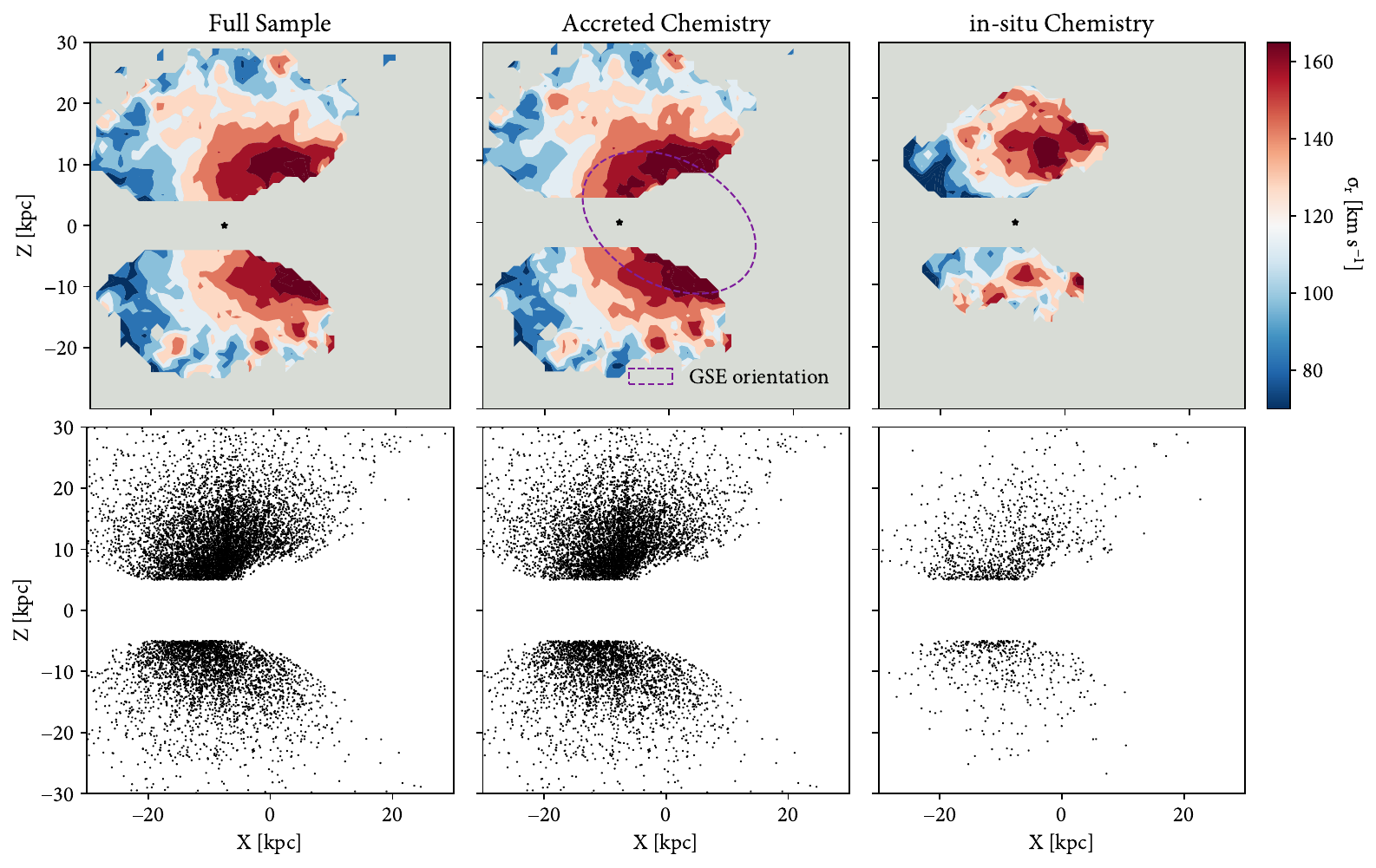}
    \caption{Galactocentric $XZ$ projection of $\sigma_r$, analogous to Figure \ref{fig:xy}. In the center panel, we overplot the iso-density contour of GSE in this projection as measured by \cite{han22b} in purple dashed ellipse. The accreted $\sigma_r$ contours are tilted with respect to the disk in a similar direction, hence breaking mirror symmetry about the Galactic plane. The in-situ $\sigma_r$ contours are also mirror asymmetric, showing preferentially higher $\sigma_r$ values above the plane. Furthermore, the in-situ $\sigma_r$ shows two distinct components: a cold structure at large radii but small $|Z|$, and a hot structure at smaller radii but large $|Z|$. We attribute the former structure to the flared thick disk, and the latter structure to the in-situ halo \citep{bonaca17, belokurov18}.}
    \label{fig:xz}
\end{figure*}

\begin{figure*}
    \centering
    \includegraphics[width=0.98\textwidth]{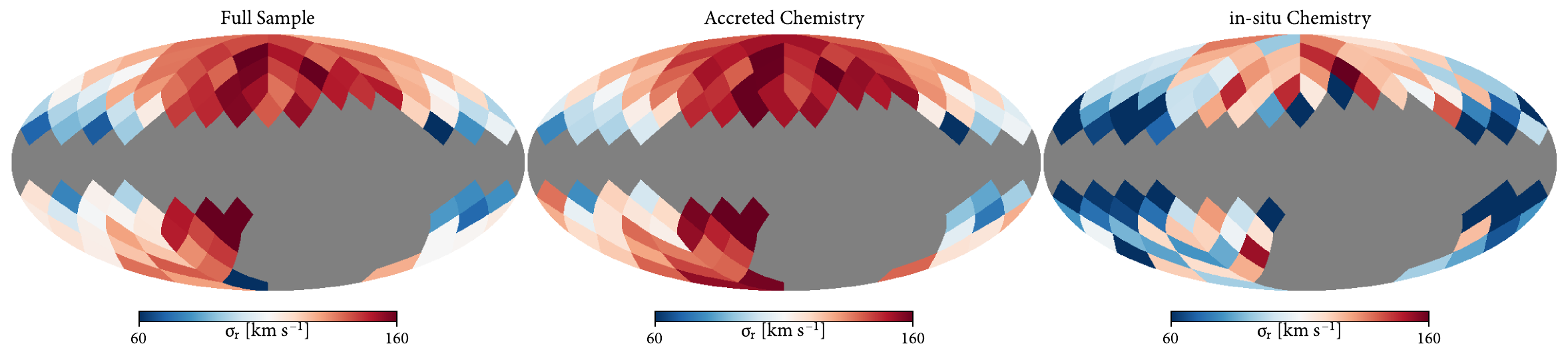}
    \caption{Galactic $(l,b)$ projection of $\sigma_r$, binned to $\texttt{nside}=4$ healpix lines of sight. Dark grey areas mark where there are less than 10 halo stars, which mostly comprise of the Galactic plane and the Southern hemisphere. In the left panel we show $\sigma_r$ of the whole halo sample, and in the center (right) panel we show the accreted (in-situ) sample. There are large variations in $\sigma_r$ in all panels. With the exception of a few lines of sight, the in-situ sample is preferentially colder than the accreted sample.}
    \label{fig:onsky}
\end{figure*}

In Figure \ref{fig:1} we show the Galactocentric velocity uncertainties of the halo sample as a function of $r_{\text{Gal}}$. This figure shows that Galactocentric radial velocity uncertainties remain roughly constant at $\sim2\text{ km}\text{ s}^{-1}$ to large radii, while tangential uncertainties increase linearly with radii beyond $30\text{ kpc}$. The dotted line marks a constant proper motion uncertainty of 0.1 $\text{mas}\text{ yr}^{-1}$ at increasing distance. The tangential velocity uncertainties converge to this line at large Galactic radii where the solar displacement from the Galactic center becomes small compared to the distance to the star. Throughout the paper, we estimate uncertainties on $\sigma_r$ and $\beta$ using the following method. For each sample, we generate $1000$ Monte Carlo (MC) realizations by sampling from the individual data errors in $V_r, V_{\phi}, V_{\theta}$, which are themselves propagated from \texttt{MINESweeper} distance, radial velocity, and \textit{Gaia} astrometric errors. We then randomly exclude 10\% of the data points in each MC sample. From the resulting 1000 measurements of $\sigma_r$ and $\beta$, we quote the median value along with $1\sigma$ and $2\sigma$ contours that contain $68\%$ and $95\%$ of the MC sample. If there are less than 10 data points in a sample to begin with, we do not report a measurement.

\section{Results}\label{sec:results}
In this section, we present measurements of $\sigma_r$ and $\beta$. We first present Galactocentric maps of $\sigma_r$ in the $XY$ and $XZ$ plane, which allows us to directly evaluate the azimuthal and meridonial symmetry of halo kinematics. We then present a heliocentric projection in Galactic $(l,b)$. Lastly, we compute spherically averaged radial profiles of $\sigma_r$ and $\beta$ in order to place our measurements in the context of previous studies.


\begin{figure}
    \centering
    \includegraphics[width=0.45\textwidth]{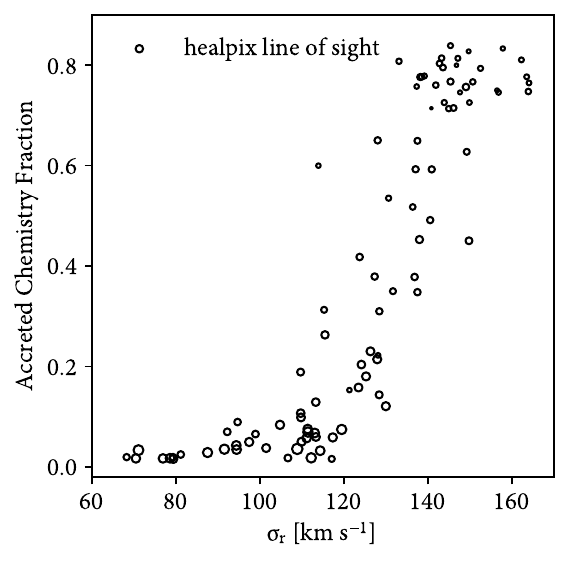}
    \caption{Line of sight averaged $\sigma_r$ versus accreted star fraction. Each circle represents an equal-area healpix line of sight from Figure \ref{fig:onsky}, and the size of the circle is proportional to the total number of halo stars. The vertical axis is computed by dividing the number of accreted stars by the total number of halo stars in a given light of sight. We find that $\sigma_r$ is as cold as 70 $\text{km}\text{ s}^{-1}$ or as hot as 160 $\text{km}\text{ s}^{-1}$ depending on the accreted fraction. The circle sizes show that the cold lines of sight do not have an anomalously low number of stars.}
    \label{fig:los_vrad}
\end{figure}

In Figures \ref{fig:xy} and \ref{fig:xz}, we map $\sigma_r$ in Galactocentric coordinates. In both figures, we define a grid from -30 to +30 kpc in each dimension in increments of 1 kpc. We measure $\sigma_r$ at each grid center within a 2 kpc radius in the given projection. We note again that $|Z|>5\text{ kpc}$ and $r_{\text{Gal}}>10\text{ kpc}$ is applied to all stars. Any region that has less than 10 stars is shaded grey. The resulting grid of $\sigma_r$ is visualized as a contour plot with 11 levels. In both figures, we show the whole halo sample in the left panel, the accreted sample in the center panel, and the in-situ sample in the right panel. Bottom panels show individual data points of each sample. For the accreted sample (center panels), we overplot measurements from previous studies of the stellar density of GSE, represented as an iso-density curve in purple dashed ellipse. The $XY$ plane measurement comes from \cite{iorio19} using \textit{Gaia} DR2 RRL, and the $XZ$ plane measurement comes from \cite{han22b} using H3 giants. The accreted $\sigma_r$ contours align remarkably well with the iso-density ellipse in both planes: both the stellar density and the kinematics of GSE are non-spherical and tilted to the disk.

\begin{figure}
    \centering
    \includegraphics[width=0.45\textwidth]{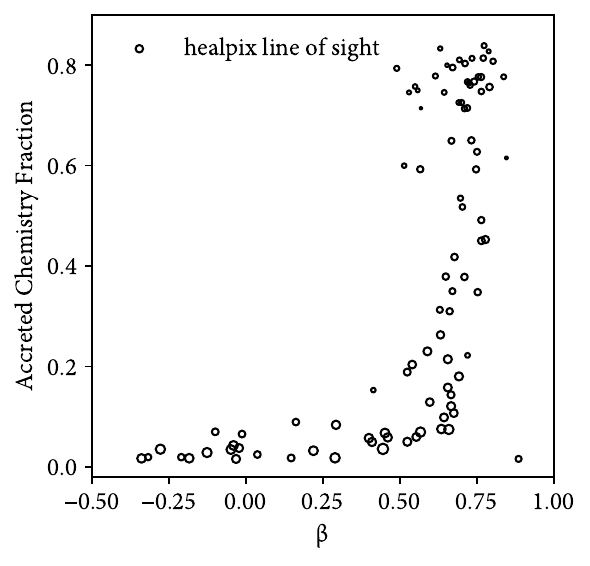}
    \caption{Line of sight averaged $\beta$ versus the accreted star fraction, analogous to Figure \ref{fig:los_vrad}. Along the lines of sight with higher accreted fraction, the average orbits of halo stars are radially biased ($\beta=1$), while those dominated by the in-situ halo are more isotropic ($\beta=0$).}
    \label{fig:los_beta}
\end{figure}

In the in-situ sample (right panels), we see a distinct cold ($\sigma_r \sim 70 \text{ km}\text{ s}^{-1}$) component at $X<-15\text{ kpc}$ and $|Z|<10\text{ kpc}$, and a hot component at $X>-8\text{ kpc}$ and $Z>10\text{ kpc}$. A plausible explanation for this peculiar configuration is that the cold component is an extension of the flared thick disk, while the hot component is the in-situ halo \citep{bonaca17, belokurov18}, an old component of the thick disk that was strongly perturbed at the time of the GSE merger. This scenario can explain the why the cold component is radially extended at cylindrical $R>10\text{ kpc}$---where the disk flare (and warp) is thought to onset---and vertically contained closer to the plane. Meanwhile, the in-situ halo is higher off of the plane and more concentrated in cylindrical radius than the flared disk ($R<10\text{ kpc}$). This radial concentration could be due to the fact that the Galactic disk was significantly smaller at the time of the merger. Another feature of the in-situ halo in Figure \ref{fig:xz} is its apparent mirror asymmetry about the Galactic plane. Future studies will investigate this interesting geometry.


In Figure \ref{fig:onsky} we divide the $(l,b)$ plane into $\texttt{nside}=4$ healpix lines of sight. In each bin we compute $\sigma_r$, and further divide the whole halo sample (left panel) into the accreted (center panel) and in-situ (right panel) components. In all panels, we see large variations in $\sigma_r$. To further explore these large variations, we plot $\sigma_r$ against the fraction of accreted stars in each line of sight in Figure \ref{fig:los_vrad}. We see a strong positive correlation in the fraction of accreted chemistry and $\sigma_r$: the halo can be as ``cold'' as 70 $\text{km}\text{ s}^{-1}$ or as ``hot'' as 160 $\text{km}\text{ s}^{-1}$ depending on how much of the sample is accreted vs. in-situ. The size of each open circle is proportional to the number of stars in the line of sight, which shows that the ``cold'' lines of sight do not have an anomalously low number of stars. In Figure \ref{fig:los_beta}, we plot the line of sight $\beta$ distribution in the same way as Figure \ref{fig:los_vrad}. Along lines of sight with larger contributions from the accreted halo, orbits tend to be more radially biased ($\beta=1$), while lines of sight dominated by the in-situ halo display more isotropic orbits ($\beta=0$). This correlation is consistent with the interpretation that the bulk of the accreted halo arose from a radial merger. Together, these figures demonstrate that spherically averaged observations of $\sigma_r$ and $\beta$ can hide significant, systematic variations on the sky. The kinematics of the Galactic halo is highly heterogeneous, and instrinsically non-spherical.

\begin{figure}
    \centering
    \includegraphics[width=0.45\textwidth]{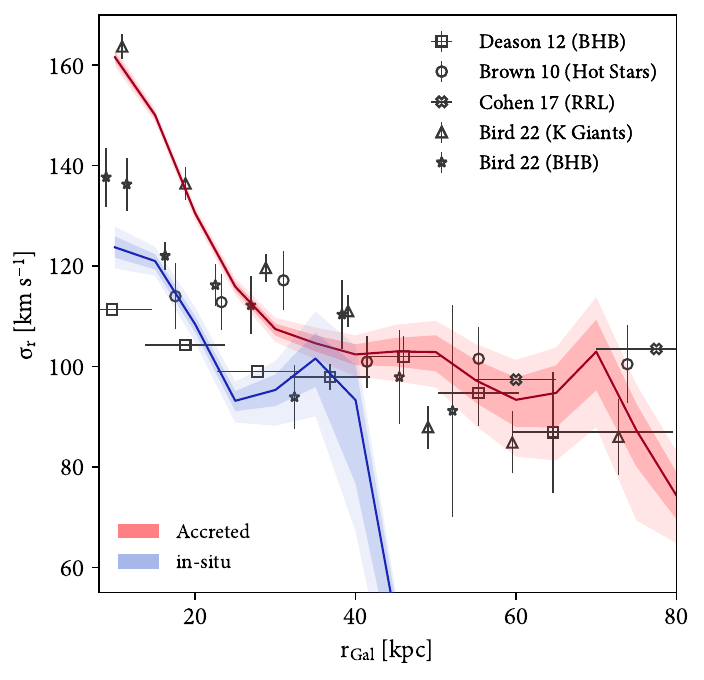}
    \caption{Spherically averaged radial profile of $\sigma_r$. We measure $\sigma_r$ in 15 linearly spaced bins between 10 and 80 kpc, and separately plot the accreted sample (red line) and in-situ sample (blue line). Shaded region denote $1\sigma$ and $2\sigma$ contours as described in Section \ref{sec:methods}. The in-situ halo notably has lower $\sigma_r$ at all radii, and its sample size drops off drastically at 40 kpc. As a result, beyond 40 kpc, most of the halo is accreted and literature values of $\sigma_r$ (open circles) are similar to each other and to our measurements. Within 40 kpc, the literature values span the full range between the in-situ and accreted profiles, likely as a result of the different selection functions used in each survey.}
    \label{fig:sigr_profile}
\end{figure}

\begin{figure*}
    \centering
    \includegraphics[width=0.9\textwidth]{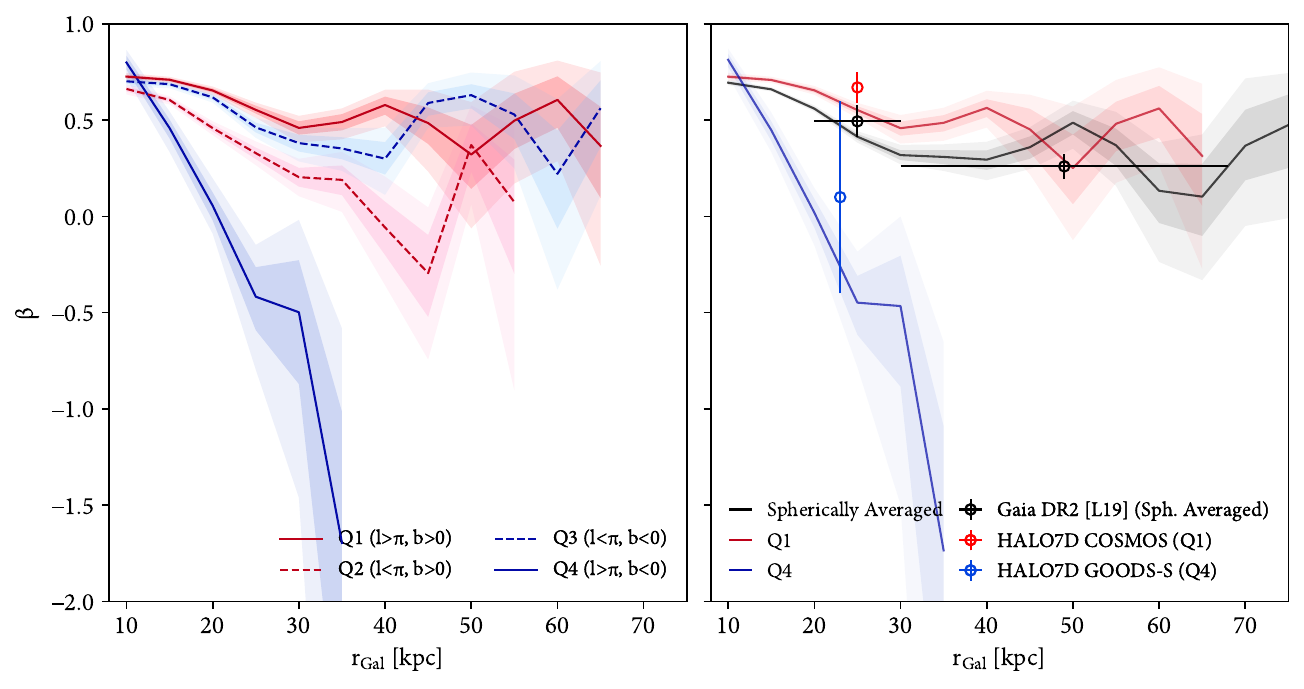}
    \caption{$\beta$ as a function of Galactocentric radius. We measure $\beta$ in 15 linearly spaced bins between 10 and 80 kpc. In the top panel, we show separate measurements in all four quadrants of Galactic $(l,b)$. In the bottom panel, we show a spherically averaged profile (black) and just the first ($l>180^{\circ}$, $b>0^{\circ}$, red) and fourth ($l>180^{\circ}$, $b<0^{\circ}$, blue) quadrants where there are HALO7D measurements \citep{cunningham19}. The black line can be compared to the \cite{lancaster19} measurements which are spherically averaged \textit{Gaia} DR2 BHBs. The red line can be compared to the HALO7D COSMOS field, which is in the first quadrant. The blue line can be compared to the HALO7D GOODS-S field, which is in the fourth quadrant. Our measurements are consistent with the literature values in both the spherically averaged case and the quadrant separated case.}
    \label{fig:beta_profile}
\end{figure*}


While these results clearly demonstrate that $\sigma_r$ and $\beta$ are not spherically distributed, it is still useful to calculate a spherically averaged radial profile in order to place our measurements in the context of previous studies. In Figure \ref{fig:sigr_profile} we plot a spherically averaged radial profile of $\sigma_r$ as measured for the accreted sample (red line) and in-situ sample (blue line) in 15 radial bins linearly spaced between 10 and 80 kpc. In open shapes we plot literature values of $\sigma_r$ spanning various tracers and radii \citep{brown10,deason12, cohen17, bird22}. Notably, the in-situ halo is colder than the accreted halo at all radii, and the sample size of the in-situ halo drops off dramatically at $r=40\text{ kpc}$. Beyond this radius, most of the halo is accreted, and literature values of $\sigma_r$ are broadly consistent with one other and also to our measurements. However, within 40 kpc there is a significant spread in the literature values of $\sigma_r$ that spans the range between our accreted sample and in-situ sample. We interpret this spread as a product of the various selection functions of each survey leading to a different ratio of accreted to in-situ stars. In addition, the anisotropic nature of the accreted halo as seen in Figures \ref{fig:xy}-\ref{fig:onsky} will further contribute to the spread in $\sigma_r$ depending on the exact lines of sights used in each survey. We note that while the BHB measurements from \cite{deason12} and \cite{brown10} seem to be systematically lower in $\sigma_r$ compared to cool giants, \cite{kafle13} separate the SEGUE BHB sample \citep{xue11} by metallicity to show that metal-poor BHBs have systematically higher $\sigma_r$ compared to metal-rich BHBs ($\sim30\text{ km}\text{ s}^{-1}$ higher at $r_{\text{Gal}}=20\text{ kpc}$), which is consistent with our result that the accreted (hence, more metal-poor) population shows a higher $\sigma_r$.

For $\beta$, we utilize the HALO7D results \citep{cunningham19} that report separate $\beta$ values for each of their fields. We thus divide the sky into four quadrants in $(l,b)$ in order to compare to literature values in isolated quadrants. In Figure \ref{fig:beta_profile} we show a spherically averaged radial profile of $\beta$ in 15 radial bins linearly spaced between 10 and 80 kpc. In the top panel, we show $\beta$ profiles in all four quadrants. In the bottom panel, we show $\beta$ profiles of the first and fourth quadrants, which can be directly compared to the HALO7D COSMOS (first quadrant) and GOODS-S (fourth quadrant) fields. We additionally show a spherically averaged $\beta$ profile in black, which can be compared to the spherically averaged profile from \cite[][L19]{lancaster19} that use \textit{Gaia} DR2 BHB stars. Across the quadrant-isolated and spherically averaged $\beta$ measurements, our results are consistent with prior studies. These figures demonstrate that the spread in literature values of $\beta$ can be explained by the nonspherical geometry of the halo kinematics, as shown in Figures \ref{fig:xy}-\ref{fig:beta_profile}.


\section{Discussion} \label{sec:discussion}

In this work, we have mapped the kinematics of giant stars across the Galactic halo using the H3 survey. Contrary to common assumption, we found that neither $\sigma_r$ nor $\beta$ are symmetrically distributed in the Galactocentric frame. Instead, the $\sigma_r$ contours shown in Figures \ref{fig:xy} and \ref{fig:xz} break azimuthal and mirror symmetry, and are aligned with the stellar density of an ancient major merger remnant \textit{Gaia}-Sausage Enceladus. This alignment bolsters the observational evidence for a non-spherical stellar halo that is tilted with respect to the Galactic disk at present day. Furthermore, the fact that the kinematics of the stellar halo are asymmetric about the Galactic plane in Figure \ref{fig:xz} is compelling evidence for a misalignment between the inner dark matter halo ($r_{\text{Gal}}<30\text{ kpc}$) and the disk \citep[e.g.,][]{han22, han23a}. We also investigated how the intrinsic asymmetries of the $\sigma_r$ distribution manifest in the Galactic $(l,b)$ coordinate system. As a result, we found large fluctuations in $\sigma_r$ across the Galactic sky: a halo of ice ($\sigma_r\sim70\text{ km }\text{s}^{-1}$) and fire ($\sigma_r\sim160\text{ km }\text{s}^{-1}$).

To place these results in the context of prior studies, we measured spherically averaged radial profiles of $\sigma_r$ and $\beta$. In Figure \ref{fig:sigr_profile} we showed that the $\sigma_r$ profiles of the in-situ halo and accreted halo bracket the literature measurements of $\sigma_r$. The spread in literature values can thus be interpreted as the consequence of a heterogeneous halo, in which the measured $\sigma_r$ is affected by the ratio of in-situ to accreted stars. This ratio is determined by the selection function of the survey, such as the specific lines of sight and metallicity biases of the tracer population. Additionally, the on-sky variations in $\sigma_r$ shown in Figure \ref{fig:onsky} can further skew measurements of $\sigma_r$ depending the exact lines of sight used in the survey. For $\beta$, we found that we can reproduce both the Galactic quadrant-specific measurements from \cite{cunningham19} and the spherically averaged measurements from \cite{lancaster19}.

All of this evidence points toward a highly non-spherical, non-axisymmetric equilibrium kinematics of the Galactic halo. In particular, the large variations in $\sigma_r$ and $\beta$ have direct consequences for spherical Jeans mass estimates. At a fixed radial profile, the Jeans mass is proportional to $\sigma_r^2$ and $\beta$, so the variation in total mass will scale as twice the variation in $\sigma_r$ and linearly to $\beta$. Clearly, the variations in $\sigma_r$ and $\beta$ seen in Figures \ref{fig:los_vrad} and \ref{fig:los_beta} are large enough to encompass the factor of two spread in literature values. Thus, in order to constrain the Milky Way mass to better than a factor of two from Jeans modeling, one needs to account for the strong intrinsic asymmetries of the halo. While the theory of triaxial equilibria was developed as early on as \cite{Schwarzschild79} and the three-dimensional solutions to the Jeans equations were presented by \cite{vandeven03}, the applications of such models to the Galaxy have been limited. The result from \cite{law10} is a notable exception, in which they find that a triaxial halo in the radial range of $20-60\text{ kpc}$ can best reconstruct the orbit of the Sagittarius stream. However, the dynamical stability of this model has been challenged by studies such as \cite{debattista13}, which find that a misalignment of the Galactic disk and the dark halo is necessary to support triaxiality. Regardless of whether or not the specific configuration of a triaxial halo is stable, it is clear that the spherical approximation can hide many of the clues that our stellar halo holds.

The effect of the Large Magellanic Cloud (LMC) on the global structure of the Galactic halo has been the focus of many studies \citep[e.g.,][]{Garavito-Camargo19,conroy21,vasiliev21,sheng24}. While the gravitational influence of the LMC is thought to take effect further out in radius ($r_{\text{Gal}}>50\text{ kpc}$), it is possible that its effects can be seen in the inner halo as well. For example, we conjecture that the mirror-asymmetry of the in-situ halo shown in Figure \ref{fig:xz} could be a consequence of the Galactic disk's reflex motion towards the LMC \citep{petersen20, chandra24}. Exploring how the equilibrium kinematics of a tilted, triaxial halo interacts with the disequilibrium effects from the LMC will be a step forward in understanding the mass distribution of the Milky Way on a deeper level.

\section{Acknowledgements}
The H3 Survey is funded in part by NSF grant NSF AST-2107253.

\bibliography{sample631}{}
\bibliographystyle{aasjournal}

\end{document}